\documentclass[aastex]{emulateapj}
\usepackage{apjfonts} 
\submitted{Accepted for publication in The Astrophysical Journal, December 15, 2003} 
\shorttitle{BLACK HOLES IN GLOBULAR CLUSTERS}
\shortauthors{KALOGERA, KING, \& RASIO }

\begin{document}

\title{Could Black-Hole X-Ray Binaries Be Detected in Globular Clusters?}

\author{V.\ Kalogera\altaffilmark{1}, A.\ R.\ King\altaffilmark{2} and F.\ A.\ Rasio\altaffilmark{1}}

\altaffiltext{1}{Northwestern University, Dept of Physics \& Astronomy, Evanston, IL; vicky, rasio@northwestern.edu}
\altaffiltext{2}{University of Leicester, Dept of Physics \& Astronomy, Leicester, UK; ark@astro.le.ac.uk}

%\affil{ $^{1}$ Northwestern University, Dept of Physics \& Astronomy, Evanston, IL\\ $^{2}$ University of Leicester, Dept of Physics \& Astronomy, Leicester, UK.\\ vicky@northwestern.edu;  ark@astro.le.ac.uk; rasio@northwestern.edu}

 \begin{abstract} 
We consider the implications of the presence of $\sim 1$ stellar-mass black hole (BH) at the center of a dense globular cluster. We show that BH X-ray binaries formed through exchange interactions are likely to have extremely low duty cycles ($<10^{-3}$), consistent with the absence of BH X-ray transients in Galactic globular clusters. In contrast, we find that BH X-ray binaries formed through tidal capture would be persistent, bright X-ray sources. Given the absence of any such
source and the very high interaction rates, we conclude that tidal capture of a main-sequence star by a BH most likely leads to the complete disruption of the star. 
 \end{abstract}


\section{INTRODUCTION}
\label{sec:intro}

Black-hole X-Ray Binaries (BH XRBs) have been reliably identified as soft X-ray transients with dynamical mass measurements of the accreting compact objects~\citep{Lasota}. More than a dozen are known at present~\citep{Orosz}, the majority of them with low-mass donors~\citep{VK99}. Their transient behavior is currently understood in the context of a viscous disk instability  including the effects of disk irradiation~\citep{KingBook}. 

%Although the general physical framework of this instability has successfully explained a number of observational characteristics, some of the details (e.g., outburst recurrence times) are not well understood and the predictability of the theory for such observables is limited. 

It has been noted for some time now that no BH XRB has ever been observed in a
Galactic globular cluster (GC), even though large numbers of neutron-star (NS) XRBs exist in GCs \citep[e.g.,][]{GrindlaySci}. 
Indeed dynamical interactions in a dense star cluster can greatly enhance formation rates for close binaries with compact objects. This is well understood theoretically~\citep{HutPASP,RPR,SP} and well established observationally~\citep{Pool}. If BH XRB formation were similarly enhanced in GCs, some BH X-ray transients should have been detected in outburst by now. Instead, all soft X-ray transients and bright, persistent X-ray sources discovered in GCs have been shown to harbor NS with the detection of Type I bursts~\citep[see][]{Jonker, Wijnands, Heinke,Rutledge,Deutsch}. 

The absence of BH XRB from GCs has been interpreted as
evidence that BHs evolve differently in clusters
\citep{KHM,SH}. Primordial BHs in GCs are throught to go
through the Spitzer mass-segregation
instability~\citep{Spitzer,Wat2000}, concentrating in a dense core and effectively decoupling from the rest
of the cluster.  Through dynamical interactions they form binaries
with two possible outcomes: (i) either the central BH cluster
completely evaporates (ejecting BHs through strong
interactions) leaving behind at most $\sim 1$ BH (single or
binary), or (ii) successive mergers of binary BHs in the cluster
driven by gravitational radiation form a $\sim
10^{2}-10^{3}$\,M$_{\odot}$ intermediate-mass BH~\citep{MH}. In the
latter case one might expect this IMBH to become visible as an
ultraluminous X-ray source (ULX), if it is not ejected from the
cluster by gravitational radiation recoil during the merger process~\citep{RR,FHH}. 
While no such source has been observed in the
old Galactic GCs, ULXs have been observed in younger extragalactic
star clusters~\citep{FW}.

In this paper, we explore the consequences of a single $\sim 10$\,M$_{\odot}$ BH remaining in the dense GC core.  We will show that the absence of any detected BH X-ray transient in a Galactic GC does not necessarily rule out the presence of $\sim 1$ BH in every cluster.

\section{DYNAMICAL FORMATION OF BH XRBs IN CLUSTERS}
 \label{sec:dynamics}

Black holes are formed from stars with initial masses $\ga 20\,M_\odot$, which
end their lives in $\la 10\,$Myr. On this timescale it is unlikely that relaxation processes
will have affected the cluster structure significantly (cf.\ G\"urkan et al.\ 2003). We make the standard assumption that kicks from asymmetric explosions scale with mass (compared to NSs) and therefore are too low ($\lesssim 50\,{\rm km}\,{\rm s}^{-1}$) to eject many BHs from the cluster~\citep[in contrast to NSs, where one faces a serious ``retention problem''; see, e.g.,][]{PRP}. 
After the evolution of massive stars down to $\sim 10\,M_\odot$, BHs become the most massive 
objects in the cluster and their subsequent dynamical evolution is driven by two-body
relaxation and mass segregation.
 
Since BHs are expected to be distributed throughout the cluster initially,
their mass segregation timescale is proportional to the initial 
half-mass relaxation time of the whole cluster, $\tau_{\rm rh}^{\rm cl}$, and inversely proportional to the BH mass $M_{\rm BH}$,
 \begin{equation} 
  \tau^{\rm BH}_{\rm seg} \simeq \frac{\left< m \right>}{M_{\rm BH}} \times \tau_{\rm rh}^{\rm cl}, 
 \end{equation} 
 where $\left< m \right>$ is the average stellar mass in the cluster~\citep{Fregeau02}. Typical values for Galactic GCs are $\tau_{\rm rh}^{\rm cl}\sim 10^{9}$\,yr and $\left< m \right>\la 1$\,M$_{\odot}$, so that a population of  $\sim 10$\,M$_\odot$ BHs is expected to decouple dynamically  in 
$\la 10^8$\,yr. 
 
The subsequent evolution of the BH sub-cluster is dominated by BH--BH interactions on their {\em own} relaxation time,
  \begin{equation} 
  \tau^{\rm BH}_{\rm r} \sim 10^6\,{\rm yr}\left(\frac{\sigma}{10\,{\rm km\,s}^{-1}}\right)^{3}\left(\frac{M_{\rm BH}}{10\,{\rm M}_\odot}\right)^{-1}\left(\frac{\rho_c}{10^5\,{\rm M}_\odot\,{\rm pc}^3}\right)^{-1},  
   \end{equation} 
   where $\sigma$ is the 3D velocity dispersion of BHs and $\rho_c$ is the mass density of the BH sub-cluster~(see eq.\ 8-71 in Binney \& Tremaine 1987; hereafter BT). Here $\sigma\sim 10\,{\rm km\,s}^{-1}$ is set by the depth of the GC potential, but $\rho_c$ is rather uncertain. Since decoupling occurs when the BH density becomes comparable to the density of the background stars, we adopt a value comparable to the central density  of typical pre-collapse GCs today~\citep{Fregeau03}.
 
BHs are ejected initially out to the GC halo (from where they sink back to the core on a short timescale $\sim 0.1\tau_{\rm rh}^{\rm cl}$ due to dynamical friction), but, as the BH sub-cluster collapse proceeds and binaries harden on $\tau^{\rm BH}_{\rm r}$, very soon they are ejected out of the GC altogether. We can estimate an upper limit to the timescale for complete evaporation by using the classical result for exponential evaporation in systems of equal point masses: 
$\tau_{\rm ev} \simeq 300\,\tau_{\rm r}$~(see eq.\ 8-79 in BT). Therefore the time for all BHs to evaporate and only $\sim 1$ to remain in the cluster is 
   \begin{equation} 
   t^{BH}_{\rm ev} \lesssim 2\times 10^9\,{\rm yr}\,\ln \left(\frac{N_{\rm BH}}{10^{3}}\right),  
   \end{equation}
 where $N_{\rm BH}$ is the initial number of BHs in the GC. In reality, evaporation is a more complex process driven mainly by inelastic collisions of binary BHs, possibly affected by gravitational radiation, and the time for complete evaporation is likely to be substantially smaller than the above classical estimate~\citep{PZMcM}. Here, we conservatively adopt $t^{BH}_{\rm ev}\sim 10^{9}$\,yr. Since this timescale is comparable to the typical relaxation time 
 $\tau_{\rm rh}^{\rm cl}$ 
of the whole cluster, we expect that, at the end of the evaporation process, the density of normal 
stars in the cluster will have started increasing through gravothermal contraction, and the 
remaining $\sim 1$ BH at the center of the cluster will be interacting significantly with normal stars. There are two interaction processes through which a BH can acquire a normal stellar companion: exchange interactions of the BH with primordial binaries~\citep{HH,SP} and tidal captures (TC) of main-sequence (MS) stars by the BH~\citep{FPR,PT,LO}. 
 
The delay between the creation of an XRB progenitor and the birth of the XRB (when mass transfer is initiated through Roche-lobe overflow) can be neglected if the BH companion is already 
somewhat evolved, i.e., its mass is comparable to the MS turn-off (TO) mass. This is 
likely to be the case for several reasons: (i) the GC center is dominated by the most 
massive stars because of mass segregation; (ii) exchange interactions have a much higher probability of 
ejecting the least massive object, so that the most massive star typically remains as a companion to 
the BH; (iii) the TC cross section is larger for more massive MS stars.
 
The characteristic timescale for the BH to acquire a binary companion  is equal to the collision time with its ``target'' binary or single star. We assume that the collision cross section is dominated by gravitational focusing, which is likely the case in all GC cores (\S\,8.4 in BT). We also 
assume that $M_{\rm BH} \gg \left< m \right>$ and we average the collision rate over a Maxwellian velocity distribution with 3D dispersion $\sigma$ for the targets.
For exchange interactions with primordial binaries the collision time is then 
 \begin{eqnarray}
 &  \tau_{\rm ex} \simeq 3\times 10^8\,{\rm yr}\, & \left(\frac{f_{b}}{0.1}\right)^{-1}\, \left(\frac{n}{10^5\,{\rm pc}^{-3}}\right)^{-1}\,\left(\frac{\sigma}{10\,{\rm km\,s}^{-1}}\right)\,
 \nonumber \\
& & \times \left(\frac{A}{1\,{\rm AU}}\right)^{-1}\,\left(\frac{M_{\rm BH}}{10\,{\rm M}_\odot}\right)^{-1},  
 \end{eqnarray} 
 where $f_{b}$ is the binary fraction in the cluster core, with stellar density $n$, and $A$ is the maximum distance of closest approach for an exchange to occur, comparable to the binary semi-major axis \citep{HH}. 
 
For tidal captures of single stars, the collision time is
 \begin{eqnarray} 
&  \tau_{\rm TC} \simeq 10^9\,{\rm yr}\, & \left(\frac{n}{10^5\,{\rm pc}^{-3}}\right)^{-1}\,\left(\frac{\sigma}{10\,{\rm km\,s}^{-1}}\right)\, \nonumber \\
  & & \times \left(\frac{r_{\rm TC}}{5\,R_{\odot}}\right)^{-1}\,\left(\frac{M_{\rm BH}}{10\,{\rm M}_\odot}\right)^{-1},  
 \end{eqnarray} 
where $r_{\rm TC}$ is the {\em maximum} closest-approach distance for the capture to occur. For a $10$\,M$_\odot$ BH and a $\lesssim 1$\,M$_\odot$ MS star, we estimate $r_{\rm TC}\lesssim 5$\,R$_\odot$ by extrapolating the results of \cite{LO}.  Note that the final fate of TC binaries is very uncertain. Previously discussed TC scenarios invovling NSs or white dwarfs have run into many difficulties.
Problems have been pointed out about the TC process
itself, which, because of strong nonlinearities in the regime relevant to GCs, is far more likely to result in a merger than in the
formation of a long-lived binary (Kumar \& Goodman 1996; McMillan et al.\ 1990; Rasio \& Shapiro 1991). Moreover, the basic predictions of these TC scenarios are at odds with many observations of binaries and recycled pulsars in GCs (Bailyn 1995; Shara et al.\ 1996).
For BHs, which are even more massive and produce an even stronger tidal perturbation, it is likely that TC does not lead to BH XRBs. Nevertheless, we take it into account for completeness. TC events with red giants (RG) can lead to BH with low-mass degenerate cores as companions on a timescale comparable to TC with MS stars (lower fraction of RG among single stars [$<10$\%] is compensated by larger $r_{\rm TC}$ [$>10$]). Such low-mass companions would be easily replaced by MS stars in a subsequent interaction and  would lead to wide binaries similar to those produced by exchange interactions.

We now turn to the evaluation of these interaction timescales (eqs.\ [4] and [5]) for typical dense GCs.
It is useful to consider three representative cases for cluster properties. 

The first case corresponds to a fairly dense, but non-core-collapsed cluster, such as 47~Tuc, 
with a central density $n\sim 10^{5}$\,pc$^{-3}$~\citep{Freire}. We adopt a binary fraction 
$f_{b}=0.1$, typical of dense GCs with resolved cores 
(Albrow et al.\ 2001). Such a cluster is thought to support itself in a state of quasi-static 
equilibrium against gravothermal contraction by primordial binary ``burning''~\citep{Fregeau03}.   
The binaries most likely to interact with a BH in the cluster core are the widest 
among the surviving (hard) systems, which have $A\sim 1$\,AU. Thus we obtain $\tau_{\rm ex} \sim 3\times 10^{8}$\,yr. This short timescale implies that 
a BH will have had multiple exchange interactions during the recent GC dynamical history
and that it would be very unlikely for a BH not to have a binary companion in this kind of environment. 
For the same cluster parameters, the TC time is $\tau_{\rm TC} \sim 10^{9}$\,yr,
implying that this process is less important than exchange interactions, but could also have happened. 

Next we consider a more extreme environment typical of a ``core-collapsed'' cluster such 
as NGC~6397, with a central density $n\sim 10^{6}$\,pc$^{-3}$. Most binaries are expected 
to have been destroyed or ejected from the cluster, so we adopt $f_{b}=0.01$ as a typical value,
consistent with observations~\citep{CB}. Since the few surviving binaries will have been hardened considerably~\citep{Fregeau03}, we take $A\sim 0.1$\,AU. 
We thus obtain $\tau_{\rm ex} \sim 3\times 10^{9}$\,yr,
while the TC time in this case is $\tau_{\rm TC} \sim 10^{8}$\,yr. Assuming that a 
star can survive TC by a BH, these results once again imply that it would be unlikely for 
the BH not to have a companion, but in this case TC would dominate over exchanges. 

Last we consider an even denser cluster similar to M15, with 
a central density $n\sim 10^{7}$\,pc$^{-3}$~\citep{Dull}. In this case no primordial binaries are expected to have survived 
in the cluster, so we assume $f_{b}=0$ and exchange interactions are unimportant (although wide binaries may still form through TC with red giants). The TC time becomes $\tau_{\rm TC} \sim 10^{7}$\,yr and implies an extremely high rate of tidal
interactions. 

\section{X-RAY DUTY CYCLES }
 \label{sec:DutyCycles}

We estimate the 
duration of any mass-transfer phase as a fraction of the timescale for replacing the BH companion 
through dynamical interactions. We also consider whether X-ray emission is persistent or transient (within the viscous disk instability model). We are then able to estimate an effective X-ray 
{\em duty cycle} ($DC$). $DC$ values close to unity, imply a high probability of 
detecting a BH XRB, since $DC=1$ would correspond X-ray emission being ``on'' throughout its lifetime, even though the BH may go through several companions.

In all three cases of GC properties considered above we find interaction 
timescales much shorter than typical GC ages ($\ga 10^{10}$\,yr). Thus recent companions are most important, and these will have masses comparable 
to the current TO mass $\simeq 0.8\,M_{\odot}$. 

Post-exchange binaries are relatively wide, with separations $A_{\rm ex}\sim 1-10\,A$ 
(where $A$ is defined
in eq.~[4]; they are also eccentric, but tidal circularization is very likely before the onset of mass transfer). This is simply from energy conservation, accounting for the higher BH mass, and in agreement with the results of detailed scattering experiments (J.~Fregeau, private communication). Therefore BH companions acquired through exchanges may not always encounter Roche lobe overflow. When they do, this will typically occur on the Red Giant Branch (RGB). For GCs like 47~Tuc, where the typical post-exchange orbital separation is $\sim 1-10$\,AU, the X-ray lifetime would be much shorter than the RGB lifetime $\tau_{\rm RGB}$. In addition, the widest post-exchange binaries may well be destroyed by another interaction (removing the  companion) before the onset of mass transfer. As an example, consider a binary with a $0.8\,M_\odot$ TO star and a $10\,M_\odot$ BH: the Roche lobe radius is $\simeq 0.2\,A_{\rm ex}$, so that no mass
transfer occurs if $A_{\rm ex} > 2\,$AU. For $A_{\rm ex}=1\,$AU, the mass-transfer duration is only $\simeq 10^7\,$yr~\citep[for a metallicity $Z=0.001$;][]{Schaller}.

In tighter systems with post-exchange separations $A_{\rm ex}\sim 0.1-1\,$AU (through interactions
with very hard binaries in a denser GC like NGC~6397), TO stars 
could fill their Roche lobes soon after the interaction. 
A strict upper limit to the duration of the mass transfer 
is then set by the RGB lifetime 
$\tau_{\rm RGB}$ of the stellar companion to the BH.
For a 0.8\,M$_{\odot}$ star, we have $\tau_{\rm RGB}\lesssim 10^{9}$\,yr. This is still shorter than the interaction timescale $\tau_{\rm ex}\simeq 3\times10^9$\,yr estimated for a cluster like NGC~6397.  
Therefore, {\em for all clusters where exchanges are important} (or post-exchange-like binaries are formed through TC with red giants), we find
$\tau_{\rm MT}/\tau_{\rm ex} \ll 1$. 

Next we consider the disk stability in post-exchange mass-transfer binaries with
orbital separations in the range $\sim 0.1-2$\,AU (orbital periods $\sim 4-400$\,d). It has been shown that such wide binaries with a mass ratio $\sim 0.1$ should be transient systems~\citep{KKB,KFKR}. Although theoretically it is difficult to estimate the transient duty cycles, observationally it appears that all systems with more than one outburst have duty cycles $f_{\rm X} < 0.01$, and for wide binaries it may be even lower~\citep{TKR}. Such low duty cycles imply that the probability of detecting a wide BH XRB as an X-ray source (during an outburst) in a GC is vanishing. Even if we adopted an optimistic average value of 
$\tau_{\rm MT}/\tau_{\rm ex} = 0.1$ for all Galactic GCs, the overall duty cycle 
$DC = f_{\rm X}\,\tau_{\rm MT}/\tau_{\rm ex} < 10^{-3}$ would be compatible with the absence of
any detection in the $\sim100$ Galactic GCs.

In contrast to post-exchange binaries, TC binaries are expected to form systems with 
very short orbital periods. Taking the typical closest-approach distance for TC to be 3\,R$_\odot$, we find the orbital period of the circularized binary to be $\sim 10$\,hr. Such tight binaries would contain small accretion disks, 
heated significantly by irradiation. For the low-mass ratios relevant to BH XRBs of interest here, 
it has been shown that the disks are stable throughout the mass transfer phase
\citep[see, e.g., middle panel of Fig.\ 1 in][]{KKB}. Therefore we would expect TC BH XRBs to 
be persistent X-ray sources. In these binaries, mass transfer is thought to be driven by magnetic 
braking initially, and gravitational radiation when the companion mass decreases below 
$\simeq 0.3$\,M$_{\odot}$. Typical X-ray lifetimes of such systems are estimated to be 
$\tau_{\rm X}\sim (1-10)\times 10^{9}$\,yr~\citep{HHKS}. Based on this estimate and the
timescales computed in the previous section we expect the effective X-ray duty cycle of 
such binaries to be $DC\sim 1$. 
At face value this result would imply that, in any dense GC containing at least one ~$\sim 10\,M_\odot$
BH, a  TC BH XRB  should have been detected as a bright, persistent X-ray source.
The absence of any such detection could be used to argue against the existence of any BH in these 
clusters. However, a more reasonable conclusion is that, although TC
of MS stars by a BH can occur at very high rates in dense GC cores, no stable binary can form through this process.

\section{DISCUSSION}
 \label{sec:discussion}

We have examined the dynamical and mass transfer evolution of BH XRBs that could possibly form in GCs, if $\sim 1$ BH remains in each cluster after $\sim 10^9\,$yr. For a wide range of cluster 
properties representative of dense Galactic GCs, we have found that (i) post-exchange BH XRBs should 
have extremely low X-ray duty cycles, but (ii) BH XRBs formed through TC should be persistent
bright X-ray sources. We concluded that TC cannot form long-lived XRBs and that our
theoretical expectations for exchange interactions are then consistent with the absence of any 
detected BH XRB in Galactic GCs. Conversely, a future detection of a BH XRB at the center of a GC
would imply that either wide binaries with low mass ratios can have disks with high X-ray
duty cycles (if the period of the observed system were at least a few days) or that TC 
can in fact lead to long lived BH XRBs (if the observed period were $\sim 10\,$hr). 

If TC of a MS star by a BH does not form a
binary, one may ask what, if any, might be the observable consequences
of such an event? For a $\sim 10 M_\odot$ BH and typical GC
velocity dispersions, we expect very little mass loss on
a hydrodynamic timescale.  Thus, if the MS star is disrupted,
essentially all its mass will remain bound to the BH, in a
rapidly rotating torus~\citep[as in collisions with a NS;
see][]{RS91,DBH}. On the hydrodynamic timescale $\sim$\,few\,hr, the gas
will settle in hydrostatic equilibrium.  On a much longer angular-momentum-transport  timescale $t_{\rm visc}$, accretion
onto the BH will commence.  Unless $t_{\rm visc}$ is extremely long
($\ga 10^6$~yr), the accretion rate $\dot M \sim 1{\rm
M}_{\odot}/t_{\rm visc}$ must be highly super-Eddington, and 
the mass  will probably be ejected
with speed $v \sim (\dot M_{\rm Edd}/\dot M)c$~\citep{KP}. For  realistic $t_{\rm visc}$, $v$ easily exceeds the GC escape velocity, leading to the complete disruption of the MS star and ejection of the gas after a very brief X-ray phase ($\sim 100$\,yr based on simple energetics for a BH accretion efficiency $\sim 0.1$). 

We have focused here on old GCs like those found in our Galaxy, where the absence of BH XRB is well established (cf.\ recent claims for BH XRBs in GCs of elliptical galaxies; Sarazin et al.\ 2003).  In younger GCs, BH companions would be more massive, possibly leading to
post-exchange binaries with persistent X-ray emission and high
detection probability. Such conditions could explain the recent {\em Chandra\/} observations of large numbers of bright X-ray sources associated with young GCs and with X-ray luminosities above the Eddington limit for a NS\citep[e.g.,][]{Macc}.

We have not considered in detail a {\em binary\/} BH system that could be retained at the end of the BH evaporation phase (see Colpi et al.\ 2003). We expect that interactions with normal stars will lead to rapid hardening of the binary. In principle, such a process can result either in a final BH merger, or in the ejection of a tight BH 
binary through classical or relativistic recoil (hence the absence of a BH XRB). It is easy to show that the merger timescale due to gravitational radiation~\citep[see eq.\ 5 in][]{KNST} becomes shorter than the collision timescale (eq.\ [5]) long before the binary can reach high enough recoil velocities ($\gtrsim 50$\,km\,s$^{-1}$). The result would be a binary merger, and the estimates presented here would still apply. 

Finally, we note that our results can be scaled to IMBH ($\sim 10^3$\,M$_\odot$). Post-exchange binaries would be extremely wide ($\gtrsim 10-100$\,AU for $A\gtrsim 0.1$\,AU), and strong hardening would be necessary for XRB formation.  TC binaries  would have circularized to periods $\sim 10$\,d, and, if they exist, they would be transients with very low duty cycles~\citep{KHIK}. 

\vspace{-0.5cm}

\acknowledgments
We acknowledge support from the Aspen Center for Physics, where this work begun. We thank K.\ Belczynski, J.\ Fregeau, T. Maccarone, and R.\ Taam for useful discussions. VK is partially supported by a  Packard Fellowship, FAR by NASA ATP Grant NAG5-12044 and a {\em Chandra\/} Theory Grant, and ARK by a Royal Society Wolfson Research Merit Award. ARK also thanks the Theoretical Astrophysics Group at Northwestern U. for visitors' support.

\end{document}